# Tubeless Siphon Flow of Room Temperature Liquid Metal

Yujie Ding[1,2], Sicong Tan[1], Xi Zhao[1,2], Yuntao Cui[1], Jianbo Tang[3], Jing Liu[1,2, 3]*

1. Technical Institute of Physics and Chemistry, Chinese Academy of Sciences,
Beijing 100190, China;
2. School of Future Technology, University of Chinese Academy of Sciences,
Beijing 100049, China;
3. Department of Biomedical Engineering, School of Medicine, Tsinghua University,
Beijing 100084, China.
*Corresponding author. E-mail: jliu@mail.ipc.ac.cn

**Abstract**

Tubeless siphon flow of liquid metal was made possible by way of sucking the fluid via a syringe pump with a rubber tube. Through adjusting the flow rate and tube diameter, the liquid column height and diameter can be changed accordingly. A theoretical model was derived to predict the flow configuration thus enabled. Adding micro copper particles evidently increased the extensional viscosity of liquid metal. With the increase of particle concentration, the enhancement of tubeless siphon becomes more remarkable. The intermetallic compound $CuGa_2$ produced by corrosion was considered as the reason for the variation of rheological property of liquid metal.

**1. Introduction**

Tubeless siphon is a special siphon effect found in non-Newtonian fluids. In this effect, the liquid is continuously sucked into a siphon tube while the tube entrance is elevated above the free liquid surface. Since the first report of this phenomenon by Fano [1], systematic researches of the tubeless siphon or Fano flow have been conducted by rheologists with considerable interest [2, 3]. It is also employed to investigate the extensional viscosity of polymeric fluids after researchers demonstrated that the tubeless siphon technique allows extensional viscosity measurement [4, 5].

The flow characteristic of liquid metal and alloys is significant for understanding many industrial processes such as metallurgy, nuclear energy, welding and thermal spraying. During last century, numerous experimental and theoretical studies have been dedicated to the flow behavior of metallic fluids. An underlying assumption adopted in most past researches is that liquid metal is a Newtonian fluid, which means that there is a linear relationship between the viscous stress and the strain rate at any point in the fluid [6]. However, recent investigations have disclosed the non-Newtonian fluid characteristic of liquid metal system. Jeyakumar et al. [7] evaluated the flow behavior and viscosity of liquid Zn, Sn, Cd and their alloys through rotational rheometry experiments. The results showed that these metallic liquids are non-Newtonian fluids featuring a shear thinning and time-dependent flow behavior. Previous studies investigated the shear flow characteristics of liquid metal with rotational rheometer. But few experimental or analytical researches have been conducted on the extensional flow of liquid metal, especially room temperature liquid metal.

Here we reported the tubeless siphon flow of liquid metal EGaIn. The effects of flow rate and tube diameter on the siphon behaviors were experimentally investigated. According to the force balance in the flow field, a theoretical model was established for pure liquid metal. Copper particles with size in micro scale were added into the metal fluid to regulate the viscosity and subsequent flow behavior.

## 2. Materials and Methods

Eutectic gallium-indium alloy $GaIn_{24.5}$ (weight percentage 24.5% In) was adopted as the experimental fluid, which has a melting point of 15.5 $^oC$. Experiments were performed at room temperature of 25 $^oC$. Thus the EGaIn remains in liquid state during all experiments. The alloy was prepared with gallium (Ga, purity 99.99%) and Indium (In, purity 99.99%) metals purchased from the Aluminum Corp. of China with melting points of 30 $^oC$ and 156 $^oC$ respectively. Compared with polymeric solution, EGaIn owns much larger density and surface tension, which are 6280 $kg/m^3$ and 624 mN/m respectively, while relatively low viscosity $1.6956 \times 10^{-3}$ Pa·s at 25 $^oC$ [8]. Copper particles with diameter in the range of 50-80 μm were added into the liquid metal to prepare liquid-metal suspension with the method developed recently by Tang et al. [9]. Through controlling the mass fraction of the particles, suspensions with mass concentrations of 2% and 4% were prepared.

A schematic of the apparatus adopted is shown in Fig. 1(a). The liquid metal $GaIn_{24.5}$ was stored in a Petri dish of 60 mm diameter and 10 mm depth. The initial volume of the liquid metal in dish was 150 ml. Rubber tubes with two different inner diameters (2 mm and 3 mm) were used to suck out the fluid. One end of the tube is fixed at the support and inserted into the liquid metal free surface with an initial depth of 5 mm. The other end of the rubber tube was connected to a 10 ml syringe mounted on a syringe pump (Longer LSP10-2A). The liquid metal was extracted from the dish with the syringe pump at constant flow rates ranging from 10 ml/min to 50 ml/min. When the flow rate was too large or small, it was hard to sustain the tubeless siphon flow. A digital camera (Canon 60D, Japan) was used to capture the image of the siphon flow. The geometry was measured in the sequence pictures by software Photron FASTCAM Viewer (error ±2 mm). Each test was repeated three times and each measurement was repeated for six times. Particle-laden liquid metal was investigated with the same apparatus.

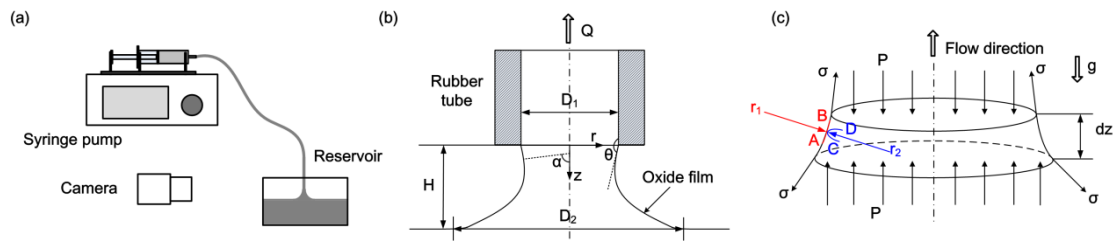

Fig. 1 (a) Schematic of the experimental apparatus; (b) Profile of the tubeless siphon flow of liquid metal; (c) Control element of liquid column.

## 3. Theoretical Analysis

As the liquid level dropped, the nozzle gradually left the liquid surface but the liquid was continuously sucked into the tube. Thus an unsupported vertical liquid column was formed between the tube nozzle and the free liquid surface. As shown in Fig. 1(b), during the continuous suction through rubber tube at volumetric flow rate $Q$, the distance $H$ between the free surface and the nozzle tip increased gradually until reaching a critical height $H_c$. The diameter of the downstream liquid column is equal to the tube inner diameter $D_1$. The upstream of the liquid column with the diameter $D_2$ is much larger than the tube diameter. A thin oxide skin formed at the outer surface and covered the liquid metal.

According to previous measurements of velocity profile in tubeless siphon flow [10, 11], the axial velocity at the centerline was much greater than the velocity of the free surface at the same axial coordinate. The flow is similar to the Poiseuille flow in a tube. However, it is difficult to measure the velocity field in opaque metal fluid. Assuming a uniform velocity profile along the z coordinate for convenience, the local velocity $U$ at each cross section is estimated as $U = Q/(\pi r^2)$, where $r$ is the section radius. Some dimensionless parameters are typically defined to characterize the fluid configuration [10, 12]. The mean diameter $D_0 = (D_1 + D_1)/2$ is used as the characteristic length. Reynolds number is defined as $\mathrm{Re} = UD_0/\nu$, where $\nu$ is the kinetic viscosity $2.7 \times 10^{-7}$ m$^2$/s, density $\rho = 6280$ kg/m$^3$, siphon velocity $U < 0.2$ m/s under current experiment conditions. Thus the Reynolds number is lower than 2300 and the flow is laminar. The capillary number is defined as $Ca = \rho \nu U/\sigma$, which represents the importance of viscous force relative to surface tension. The surface tension of EGaIn with oxide skin is $\sigma = 0.624$ N/m. Thus the capillary number is quite small ($10^{-4}$) and viscus force is negligible.

During the tubeless siphon experiments of the liquid metal, the flow field is unsteady in the Lagrangian system. The liquid column gradually deformed and the curvature of the free surface varied even though the flow rate was constant. When a large amount of liquid metal in a big vessel was used in the siphon experiments, the liquid metal level decreased very slowly and the siphon liquid column could be approximated as static with a steady configuration. According to the geometric structure in Fig. 1(b), the relation between the radial coordinate $r$ and the angle $\alpha$ is

$$\frac{dr}{dz} = \tan \alpha \qquad (1)$$

where $\alpha$ is the divergence angle between the symmetry axis and the generatrix normal.

Figure 1(c) illustrates the forces acting on the control element including surface tension on liquid/gas interfaces, gravity and pressure acting on the cross section. The momentum flux is $M = \pi R^2 \rho U^2 = \rho Q^2/(\pi r^2)$. The momentum variation in the direction along z coordinate is

$$\frac{dM}{dz} = -\frac{d}{dz}\left(\pi r^2 P + 2\pi r \sigma \cos \alpha\right) - \pi r^2 \rho g \qquad (2)$$

The pressure drop at the liquid/gas interfaces could be obtained from the Young-Laplace equation, i.e.

$$\Delta P = \sigma \left(\frac{1}{r_1} + \frac{1}{r_2}\right) = \sigma \frac{d(r \sin \alpha)}{r\, dr} \qquad (3)$$

where the $r_1$ and $r_2$ are the curvature radii of the outlines A-B and C-D.

Therefore, the gauge pressure in the siphon flow can be estimated as

$$P = \sigma\left(\frac{\sin\alpha}{r} + \cos\alpha\frac{d\alpha}{dr}\right) \quad (4)$$

The length is nondimensionlized with respect to capillary length $l_c = \sqrt{\sigma/(\rho g)} = 3.2$ mm. Therefore, the dimensionless form of the Eq. (1) and (2) are

$$\frac{d\bar{r}}{d\bar{z}} = \tan\alpha \quad (5)$$

$$\frac{\bar{r}^2}{\tan\alpha}\left[\cos\alpha\frac{d^2\alpha}{d\bar{z}^2} - \sin\alpha\left(\frac{d\alpha}{d\bar{z}}\right)^2\right] + (3\cos\alpha - 2\sin\alpha)\bar{r}\frac{d\alpha}{d\bar{z}} + \bar{r}^2$$
$$+ (\sin\alpha + 2\cos\alpha)\tan\alpha - 2We\tan\alpha = 0 \quad (6)$$

The Weber number in the above equation is given by $We = \rho U^2 r / \sigma = \rho Q^2 / (\pi^2 \sigma r^3)$, which measures the relative importance of inertia compared to surface tension. Under the given flow rate $Q$ and tube size $D_1$, the parameters $\bar{r}(\bar{z})$, $\alpha(\bar{z})$ are to be calculated. Runge-Kutta method was adopted to solve these equations with the boundary condition at the tube nozzle $\bar{z}=0$, $\bar{r} = R_1/l_c$, $\alpha=\theta$, where $R_1$ is the tube inner radius and $\theta = 160^o$ is the contact angle of liquid metal on tube nozzle.

## 4. Results and Discussions

4.1 Tubeless siphon flow of pure liquid metal and particle-laden liquid metal

The effects of siphon speed and tube diameter were investigated in a series of experiments with pure and particle-laden liquid metal. A liquid column with height of about 3 mm was gradually established after the liquid surface dropped below the nozzle tip (t=0 s), as illustrated in Fig. 2(a) (t=4 s). Due to the breaking of the force balance, the liquid metal column reached the critical height and got necked. Finally, the liquid column broke away from the nozzle and the siphon stopped. The whole siphon process lasted about 8 s. The results showed high degree of repeatability of liquid metal tubeless siphon effect.

For the test fluid with 2% suspended particles (Fig. 2(b)), the tubeless siphon effect was obviously enhanced. The maximum height of the liquid column was increased. The lifetime of the siphon effect extended to 9 s. When the mass fraction of the micro copper particles reached 4%, the suspension fluid became dim gray and rough. The tubeless siphon experiments demonstrated that this suspension fluid exhibited strong viscoelasticity. The liquid metal near the column based was sucked away and a pit was formed. Unlike pure metal fluid, the pit in the liquid metal base would not smoothen after a long time (>10 min).

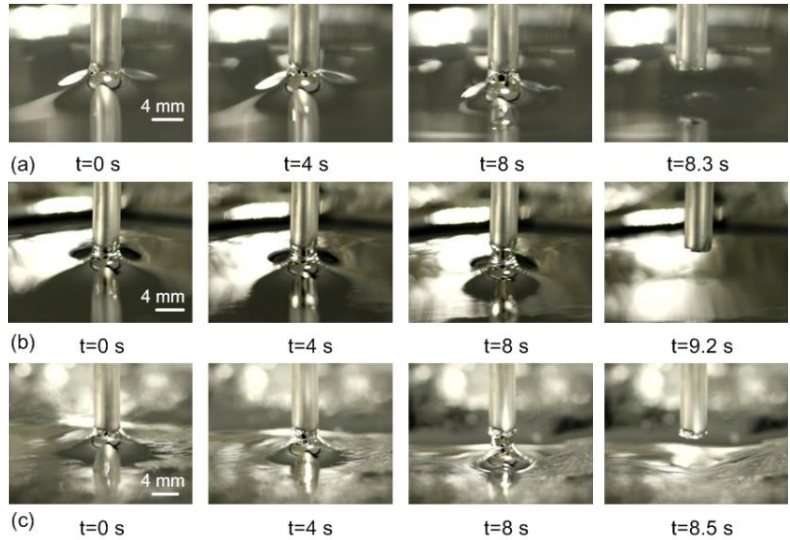

Fig. 2 Tubeless siphon effect of liquid metal fluids under 20 ml/min flow rate in tube of 3 mm inner diameter: (a) pure EGaIn; (b) particle mass fraction 2%; (c) particle mass fraction 4%.

The measurement results indicated that the tubeless siphon flow is strongly influenced by the flow rate, tube size and mass fraction of micro-particles. In Fig. 3(a), the column height of pure EGaIn decreased from 3 mm to 2 mm with the increase of flow rate. Through adding micro copper particles into the liquid metal, the critical height of liquid metal suspension was evidently enhanced, especially under high flow rate. In the experiments with a thinner tube with 2 mm inner diameter (Fig. 3(b)), the tubeless siphon flow of liquid metal showed different trends. A thinner tube benefited the formation of the tubeless siphon of pure liquid metal. The critical column heights of pure metal were improved under all flow rates. In contrast, the flow of the particle-laden liquid metals was constrained. Under higher flow rates, the vertical columns of liquid metal suspensions were unstable and short. Correspondingly, the upstream radius was also affected by the flow rate and tube size, as shown in Fig. 3(c) and (d). With the increase of the flow rate for 3 mm tube, the liquid column radius for pure metal and 2% metal suspension increased while that for 4% metal suspension decreased. However, the liquid column radius for pure metal decreased with flow rate in 2 mm tube while that for 2% and 4% metal suspensions increased.

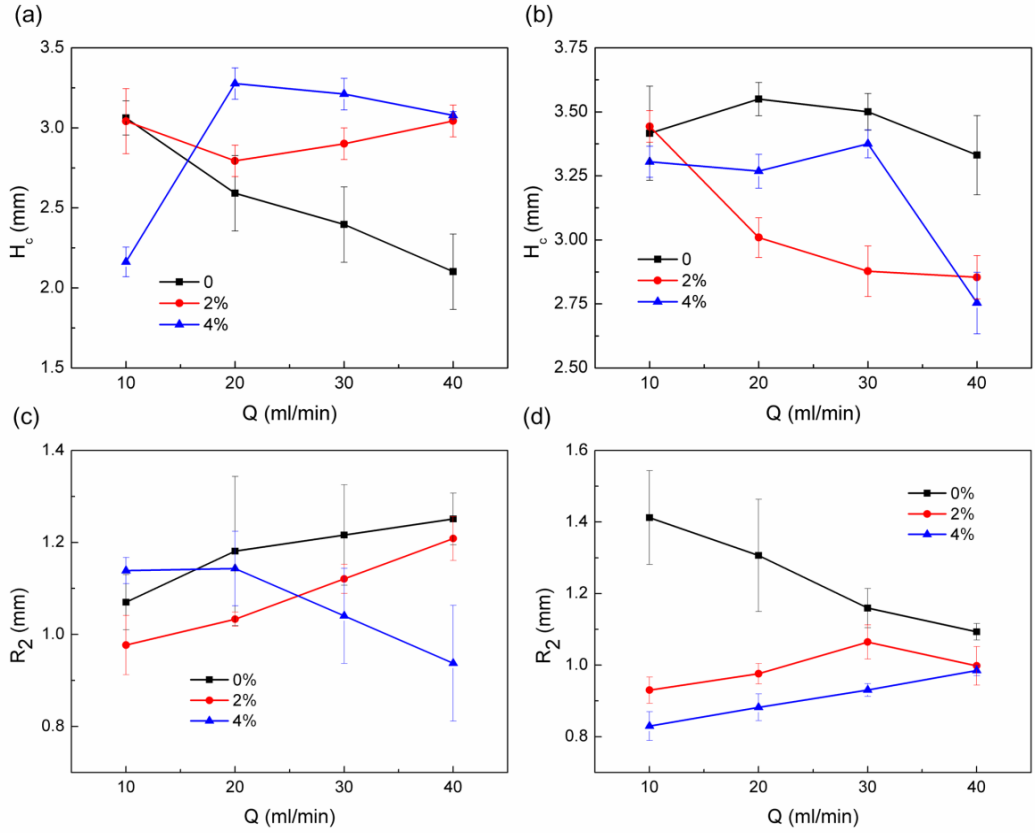

Fig. 3 Critical heights of tubeless siphon column versus flow rate in the tubes under different particle concentrations with (a) 3 mm inner diameter and (b) 2 mm inner diameter; Corresponding upstream radius of liquid metal column versus flow rate in the tubes with (c) 3 mm inner diameter and (d) 2 mm inner diameter.

4.2 validation of the model for pure liquid metal

The calculation results of tubeless siphon flow of pure liquid metal when the liquid column reaches the critical height are illustrated in the Fig. 4. The dimensionless radius increased and the total height decreased with flow rate in Fig. 4(a). On the contrary, the dimensionless radius decreased and the total height increased with flow rate in Fig. 4(b). In other words, with the increase of flow rate $Q$, the liquid column grew thicker and shorter for 3 mm tube and grew thinner and longer for 2 mm tube. This result is consistent with the experimental measurements in Fig. 3. As illustrated in Fig. 4(c) and (d), the vertical flow velocity for 2 mm tube and 3 mm tube all increased with the flow rates, while the fluid flowed faster with 2 mm tube than that with 3 mm tube.

The points in Fig. 4(a) and (b) represent the experimental data at dimensionless height $\bar{H}=0.5$ and $\bar{H}_c$. The calculated curves are close to the experimental data points in Fig. 4(a). But there is a relatively large deviation between the curves and data points in Fig. 4(b). In addition, the slope of the curves in Fig. 4(b) gradually decreased along the z coordinate, which obviously differs from practical situation. The deviation may be caused by the assumption that the vertical velocity in the fluid column is homogenous at any cross section. As illustrated in Fig. 4(c) and (d), the vertical flow velocity for 2 mm tube is much larger than that for 3 mm tube under all flow rates. Therefore, the shear effect will be more important in the thinner tube which leads to ineligible velocity gradient along the radial coordinate.

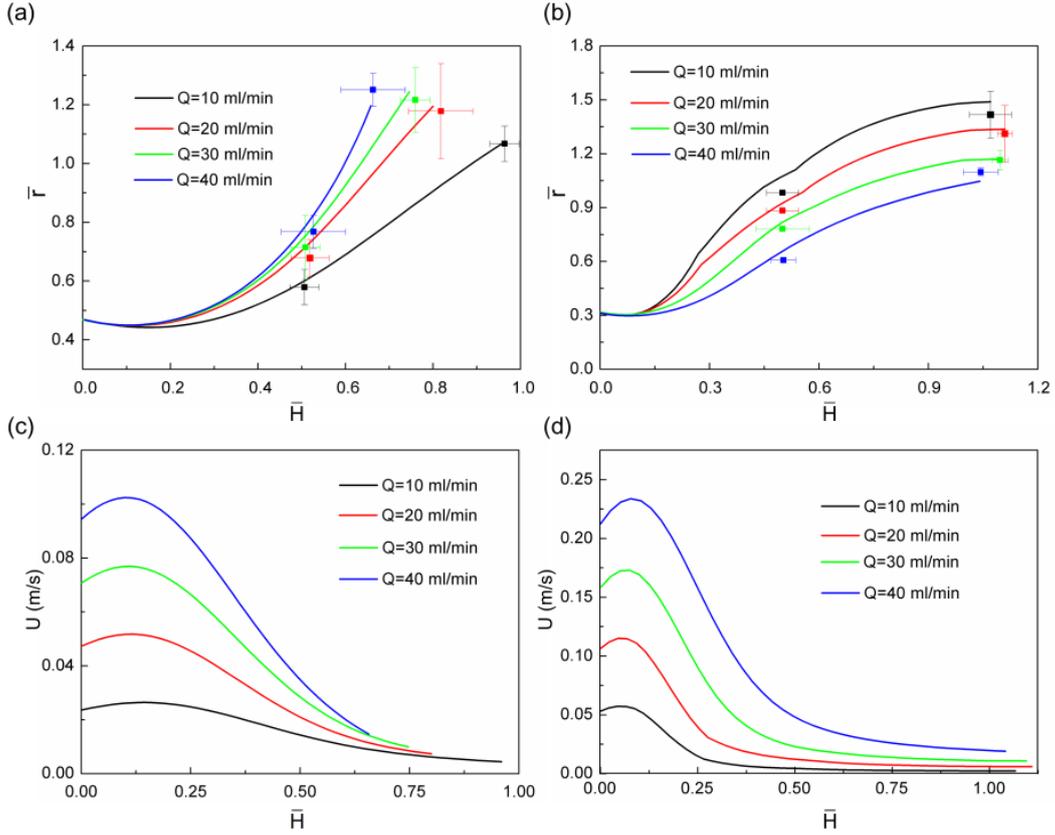

Fig. 4 Calculated configuration of tubeless siphon column of pure liquid metal with (a) 3 mm tube and (b) 2 mm inner tube under different flow rate; Velocity profile in the liquid column along the flow direction with (c) 3 mm tube and (d) 2 mm tube.

4.3 Mechanism of the particle-laden liquid metal

In order to figure out the effect of micro particles on tubeless siphon of liquid metal, some explanations are proposed and verified by experiments. When the liquid column is established, the weight of the fluid between the nozzle and the free liquid surface is sustained by the normal stress $\sigma_{zz}$ in the vertical direction, which is induced by the velocity gradient in the flow direction i.e. $\sigma_{zz} = \eta_e dU/dz$, where $\eta_e$ is extensional viscosity. The extensional viscosity of polymer solution is attributed to the orientation and extension of the polymer chain in the flow direction. According to the experiments of Wang and Joseph [13], the extensional viscosity of the polymer solutions could be greatly increased by adding sub-millimeters resin particles at moderate volume concentration. Some mechanisms were proposed based on the characteristics of the flow liquid column and particle chains. In organic suspensions and emulsions, the main forces can be of three kinds: van der Waals attraction, electrostatic repulsion due to the interactions of the electrical double layers around the particles, and the repulsion due to the interaction of absorbed large molecules on particles [14, 15, 16]. In current suspensions of copper particles in EGaIn liquid metal, there may be some different mechanisms. Considering the corrosivity of gallium-based alloy to copper, it is proposed that the interactions induced by the corrosion of copper particles lead to the increase of extensional viscosity. Figure 5 displays the SEM and XRD patterns of a T2 copper plate corroded by $GaIn_{24.5}$ alloy at room temperature for 12 h. As shown in Fig. 5(a), the copper sample becomes very coarse after corrosion and there are some corrosion products on the surface. The EDS analysis shows that the copper molar concentration is 36.66% and gallium molar concentration is 63.34%. To identify the corrosion product covering the sample surface, XRD is used to determine its phase composition. As illustrated in Fig. 5(b), diffraction peaks corresponding to Cu appear on both lines at the same place. There are also $CuGa_2$ diffraction peaks on the pattern of corroded sample. Based on the above results, it is demonstrated that a corrosion layer of intermetallic compound $CuGa_2$ gradually formed and covered the particles after the addition of copper particles into $GaIn_{24.5}$ liquid metal. Presumably, the metallic bond actions between the Cu particles and liquid metal caused the enhancement of the extensional viscosity. With the increase of the particle concentration, the effect of particles on the extensional viscosity was more evident.

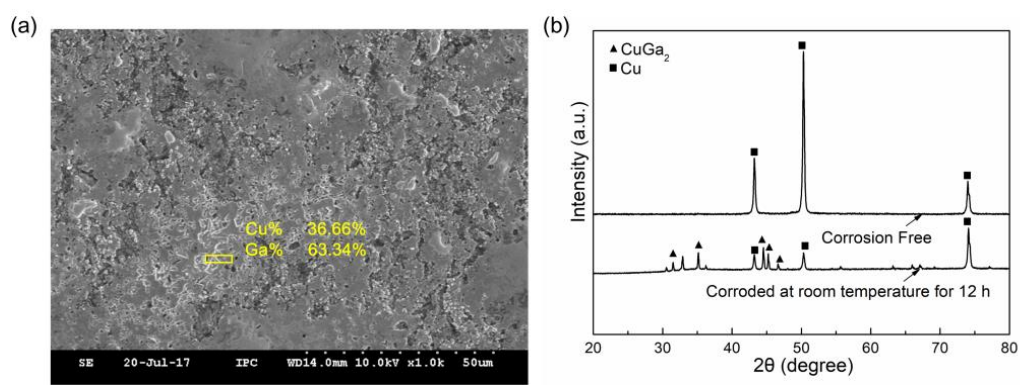

Fig. 5 (a) Morphology of the T2 copper alloy surface corroded by $GaIn_{24.5}$ alloy at room temperature for 12 h; (b) XRD analysis of corrosion-free/corroded copper surface.

## 5. Conclusion

In summary, the tubeless siphon effect of liquid metal has been investigated experimentally. A vertical free liquid column can be established with a height of about 3 mm and sustained for about 8 s under current conditions. The size of the siphon tube and flow rate has remarkable influence on the liquid column. A theoretical model is developed to predict the configuration of the liquid column. There is possibly an appropriate range of tube size and flow rate for which the tubeless siphon is easier to appear. Through adding micro copper particles into the liquid metal, the tubeless siphon effect would be enhanced. With the increase of the particle mass concentration, the effect of particles on the extensional viscosity is more obvious. SEM and XRD are adopted to determine the corrosion product on copper surface. Intermetallic compound $CuGa_2$ is found and considered as the reason for the increase of extensional viscosity. This novel phenomenon of liquid metal presents a class of characteristics of metal fluids and offers a strategy to measure the extensional viscosity of liquid metal.

**Nomenclature**

$Q$=flow rate (ml/min)

$H$=liquid column height (mm)

$H_c$=critical column height (mm)

$D_1$=tube inner diameter (mm)

$R_1$=tube inner radius (mm)

$D_2$=column base diameter (mm)

$R_2$=column base radius (mm)

$U$=local velocity (m/s)

$r$=section radius (mm)

$D_0$=mean diameter (mm)

$Re$=Reynolds number

$\sigma$=surface tension (N/m)

$v$=kinetic viscosity (m$^2$/s)

$\rho$=density (kg/m$^3$)

$Ca$=capillary number

$M$=momentum flux (kg m/s$^2$)

$\alpha$=divergence angle ($^o$)

$z$=vertical coordinate (mm)

$P$=pressure (Pa)

$r_1$=curvature radii 1 (mm)

$r_2$= curvature radii 2 (mm)

$l_c$=capillary length (mm)

$We$=weber number

$\theta$=contact angle ($^o$)

$\sigma_{zz}$=normal stress (kg/s$^2$/m)

$\eta_e$=extensional viscosity (kg/s/m)